\documentclass[useAMS,usenatbib,usegraphicx]{mn2e}

\title[Imaging and integral field spectroscopy of shocked
  $\mathrm{H_2}$ around G25.65$+$1.05]{Imaging and integral field spectroscopy of shocked
  $\mathrm{H_2}$ around G25.65$+$1.05}
\author[S. P. Todd and S. K. Ramsay Howat]{Stephen P. Todd$^{1,2}$\thanks{E-mail:
spt@roe.ac.uk (SPT)} and Suzanne K. Ramsay Howat$^{1}$\thanks{E-mail:
skr@roe.ac.uk (SKRH)}\\
$^{1}$UK Astronomy Technology Centre, Royal Observatory, Blackford Hill, EH9 3HJ, UK. \\
$^{2}$Institute of Astronomy, University of Edinburgh, Royal
Observatory, Blackford Hill, EH9 3HJ, UK.}

\begin{document}

\date{Accepted . Received ; in original form }

\pagerange{\pageref{firstpage}--\pageref{lastpage}} \pubyear{2002}

\maketitle

\label{firstpage}

\begin{abstract}

Near-infrared imaging of the emission from molecular hydrogen is a
powerful method for discovering outflows in star-forming regions.
We present new near-infrared images, long slit and integral field
spectroscopy of the ultra-compact \textsc{H~ii} region
G25.65+1.05. These new observations reveal shocked ${\mathrm H_2}$
emission associated with a bipolar outflow from a young high mass
star at the centre of the source. The physical parameters of the
outflow are discussed and compared with outflows from lower mass
stars.

\end{abstract}


\section{Introduction}

A number of recent surveys have invigorated the study of high mass
star formation by providing data on coherent samples of candidate high
mass young stellar objects (HMYSOs) (e.g. \citealt{palla91},
\citealt{molinari96}, \citealt{molinari98}, \citealt{ridge01},
\citealt{sridharan02}). Until these surveys became available the
relative scarcity of HMYSOs, the large distances to the closest
examples (a few kpc) and the high extinction to these objects had
limited their study to a few, well-known examples. Thus, there are
many questions still remaining as to the nature of high-mass star
formation, and the similarities and differences in high and low mass
stellar evolution.

Outflows are associated with the formation of stars of all masses
and are the subject of a number of the HMYSO surveys
(\citealt{shep2}, \citealt{shep1}, \citealt{zhang01}). While the
role of outflows in the formation of low-mass stars is
comparatively well studied (see for example the review by
\citealt{richer00}), far less is known about the properties and
role of outflows associated with the formation of high-mass stars
(M greater than $\sim$~8~M$_{\odot}$). A radio survey of molecular
line emission from high-mass star forming regions showed that
high-velocity molecular gas (CO) is associated with around 90\% of
these regions (\citealt{shep1}). Two sources (G25.65+1.05 and
G240.31+0.07) were mapped at higher spatial resolution and found
to have bipolar outflows. In a follow-up survey, \citet{shep2}
mapped a further ten of the best candidates for having bipolar
molecular outflows and confirmed the presence of such an outflow
in 5 of those sources. \citealt{zhang01} observed 69 candidate
high mass protostars in the CO J=2-1 transition and argue that as
many as 90\% of the sources may have outflows. \citet{beuther02a},
mapping at a higher spatial resolution, found evidence of bipolar
outflows in 21 of their 26 sources suggesting that bipolar
outflows may be associated with most young high mass stars. The first
report of molecular hydrogen line emission associated with high mass
star formation was from \citet{lee01}, following up a search for
the near-infrared counterparts of ultra-compact H~{\sc ii} regions by
\citet{walsh99}. \citet{lee01} found evidence for outflows in
IRAS~15278-5620 and IRAS~16076-5134.

These surveys confirm the importance of outflows for HMYSO
evolution. Further work is required to determine whether the
physics of the HMYSO outflows is the same as for low mass YSOs.
The observed outflows appear to have much lower collimation factor
-- between 1 and 1.8~-- than those seen from low-mass stars which
often have a collimation factor of around 10 (see, for example,
the interferometric observations of \citet{richer00}). This would
be hard to explain if the outflows are formed by the same jet
entrainment model as that believed to describe the outflows from
low-mass stars.  However, \citet{beuther02a} argued that the
observed degree of collimation could be significantly reduced by
the low spatial resolution of the maps, which could be consistent
with well collimated high-mass flows. Interferometric observations
have shown collimation factors as high as 10 \citep{beuther02b} in
flows from high-mass stars and have revealed that some of these
apparently uncollimated flows can be resolved into several well
collimated flows from separate young stars. Detection of multiple
flows in a region in which high-mass stars are forming would not
be surprising. High-mass stars are known to form in dense clusters
\citep{garay99}, so high spatial resolution observations are
essential for identifying the source of an individual outflow. The
presence of a collimated outflow would imply the presence of a
stable accretion disk and hence strengthen the view that high mass
stars are formed by steady accretion, in a similar way to low mass
stars, rather than by merging of intermediate-mass protostars in
the centre of dense clusters.

We have embarked upon a near-infrared survey of HMYSOs, which will
be presented in full in Varricatt et al. (2005, in preparation).
The survey is designed to reveal outflows by mapping at
sub-arcsecond spatial resolution in the v=1-0 S(1) line of
molecular hydrogen. NIR observations of $\mathrm H_2$ have
frequently been used to study outflows from low mass YSOs. In this
paper we present near-IR imaging, integral-field and long-slit
spectroscopic observations of new outflows in the region of the
ultra-compact \textsc{H~ii} region G25.65$+$1.05 .

G25.65$+$1.05 (also IRAS 18316$-$0602 or RAFGL7009S) is an
irregular, compact radio source, first identified at 3.6cm by
\citet{kurtz94}. It is located at a distance of 3.2kpc
\citep{molinari96}. The radio peak is coincident with an
unresolved infrared source, identified as a young B1V star with a
large K-band excess \citep{zavagno02} and is also closely
associated with methanol \citep{molinari96} and ammonia maser
emission (\citealt{walsh03}, \citealt{szymczak00}). Submillimetre
continuum observations at 350$\mu$m \citep{hunter00}, 450$\mu$m
and 850$\mu$m \citep{walsh03} are all peaked at the position of
the radio and maser sources. Observation of the CS (2-1) line by
\citet{bronfman96} shows an excellent match of the observed radial
velocity from the masers (40.8-42.4~kms$^{-1}$, \citealt{walsh03})
and the line emission (41.4~kms$^{-1}$) indicating a strong link
between the dense gas, the maser sources and the massive star.
\citealt{zavagno02} propose that one explanation for the observed
$K$-band excess from the central source may be the presence of a
disk. ISO spectroscopy of this source shows a rich spectrum of ice
features including absorption features attributed to H$_2$O,
CH$_3$OH, CO$_2$,$^{13}$CO$_2$,CO, OCS, HCOOH, HCOO$^-$,
CH$_3$HCO, CH$_4$, NH$_3$ and Silicate (see \citealt{gibb04} and
references therein). Laboratory spectra fitted to the ISO
observations suggests that the ice features arise in dense
material with temperatures in the range 10K-100K.

\section{Observations}

\subsection{Imaging}

\begin{figure}
  \centering
    \includegraphics[width=70mm, angle=-90]{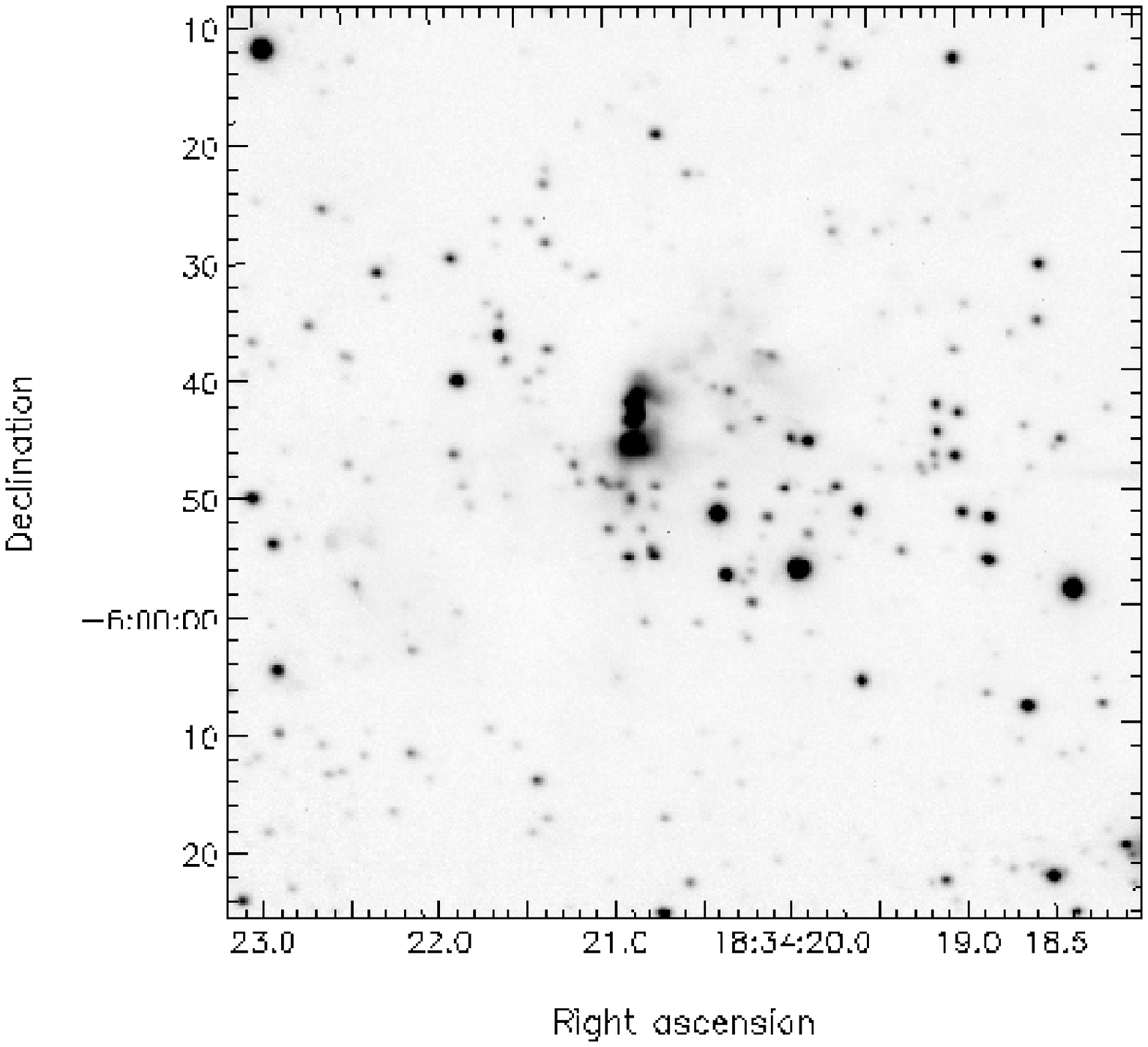}\vspace{0.6cm}\\
    \includegraphics[width=70mm,angle=-90]{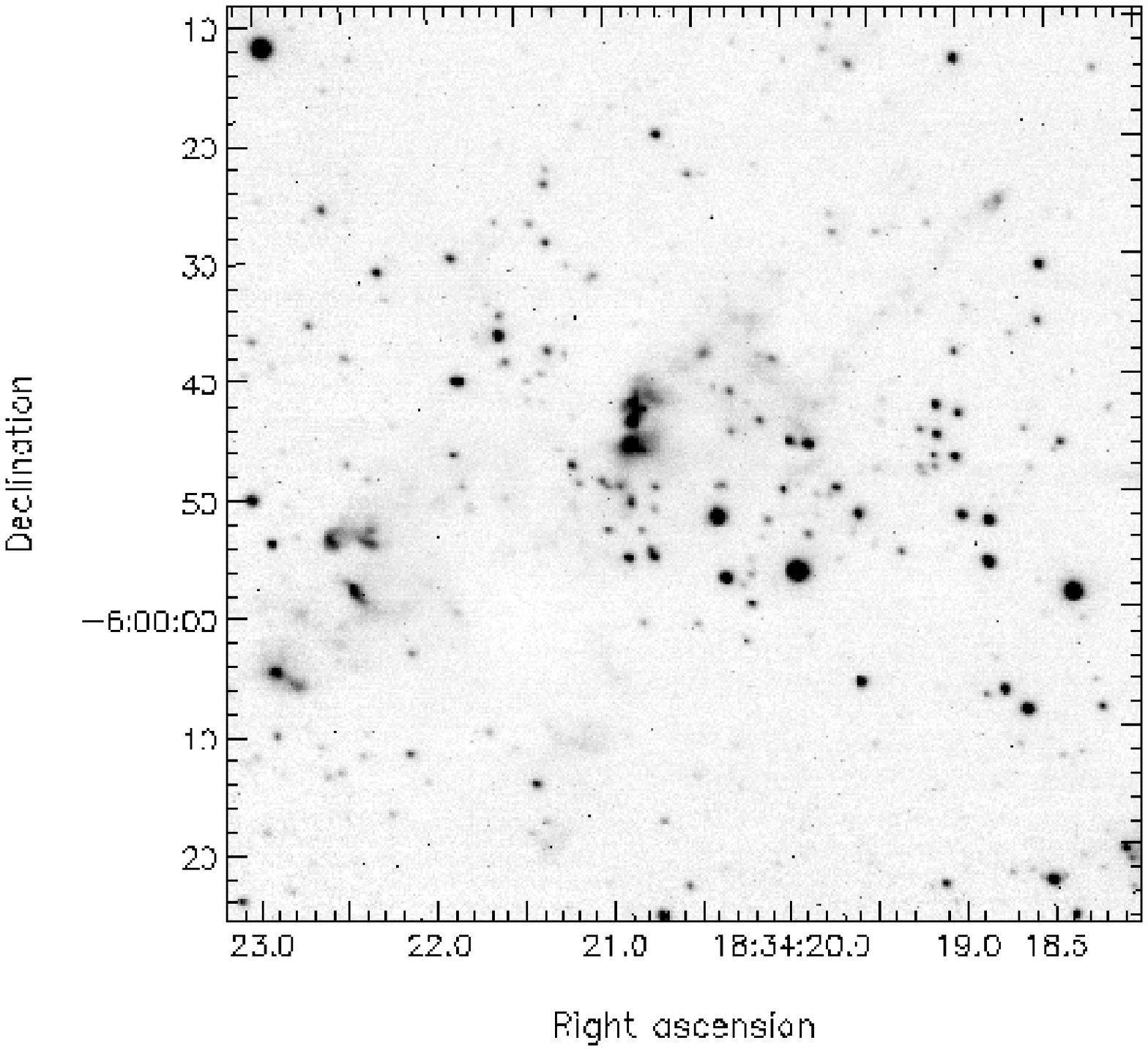}\vspace{0.6cm}\\
    \caption{Part of the UFTI mosaic images (a) K98 broad-band filter;
        (b) $1-0~\mathrm{S}(1)$ narrow-band filter. All coordinates are J2000. }
\label{ufti_images}
\end{figure}

Imaging observations were made on 29 June 2002 using the facility
infrared imager, UFTI \citep{roche03}, at the UK Infrared
Telescope (UKIRT). UFTI is a 1 to 2.5~\micron\ camera using a
$1024 \times 1024$ HgCdTe array with a plate scale of 0.091
arcsec/pixel. A single 9-point jitter in a $3 \times 3$ pattern
with offsets of 10~arcsec was observed with 60~sec exposures using
the $K98$ broad-band filter, giving 9~min on source. The same
9-point jitter pattern was repeated three times using exposures of
100~s with the 2.122~\micron\ $1-0~\mathrm{S}(1)$ $\mathrm{H_2}$
narrow-band filter giving a total of 45~min on source. The
individual frames from each filter were flat-fielded and mosaiced
together automatically by the \textsc{orac-dr} pipeline.

Sources emitting in the $\mathrm{H_2}$ line will be detected in
both the broad-band and narrow-band images but will appear
brighter relative to the continuum sources in the narrow-band
images. The two mosaics are shown in Figure~1. The narrow-band
image with 0.6~arcsec seeing was smoothed to match the 0.7~arcsec
seeing of the broad-band image. The broad-band image was scaled in
intensity using the flux from the field stars and subtracted from
the narrow-band image to leave the $\mathrm{H_2}$ emission as
shown in Figure~2. The \textsc{H~ii} region itself and some stars
around it appear negative in this subtracted image due to
differential extinction to the core compared with the field stars
used for flux scaling. The increased reddenening makes it
relatively brighter in the broad-band image. A number of point
sources are visible in addition to the extended emission. These
are ghost images of bright stars produced by the narrow-band
filter and residuals left where stars have been imperfectly
subtracted. The complete mosaic covers a $1.8 \times 1.8$~arcmin
field centred on the \textsc{H~ii} region. No $\mathrm{H_2}$
emission was detected outside the region shown here. The brightest
$\mathrm{H_2}$ emission was detected to the south-east of the
\textsc{H~ii} region, taking the form of a region of faint,
diffuse emission containing a number of bright, compact sources
(A--D). To the north-west of the \textsc{H~ii} region there is a
straight, narrow line of faint emission with a bright source (E)
at one end. There is also faint, diffuse emission to the south and
north-east of the \textsc{H~ii} region.
\begin{figure*}
\begin{minipage}{130mm}
  \centering
    \includegraphics[width=120mm,angle=-90]{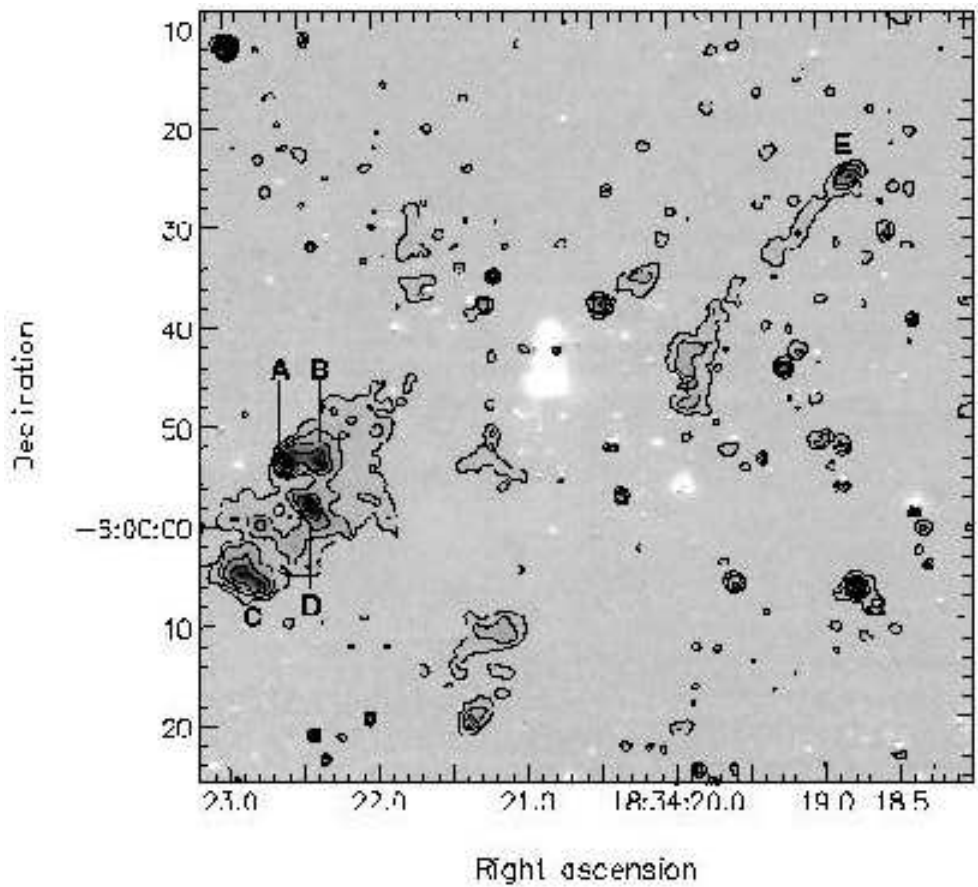}
    \caption{A continuum subtracted $1-0~\mathrm{S}(1)$ image
    created by subtracting Figure~1a from
    Figure~1b. The contours are drawn from a
    version of the image smoothed with a 0.5~arcsec gaussian and
    are logarithmically spaced with each contour a factor of two
    higher in flux than the previous one from an arbitrary lowest
    contour.  All coordinates are J2000. }
\label{ufti_images2}
\end{minipage}
\end{figure*}

\begin{figure}
  \centering
   \includegraphics[width=7cm]{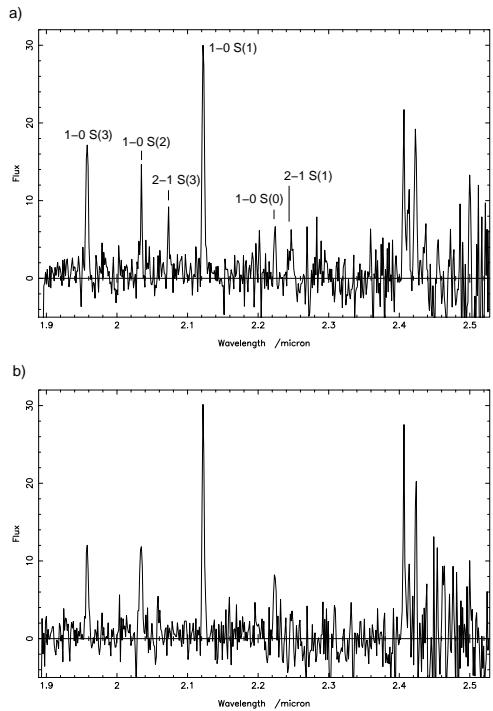}\\
  \caption{CGS4 long-slit spectra of a) source C and b) source D. The flux is in arbitrary units.}
\label{cgs4}
\end{figure}
\subsection{Long-slit spectroscopy}

Long slit spectra of sources C and D were obtained on 30 June 2002
at UKIRT using Cooled Grating Spectrometer 4 \citep{mountain90}.
The 40~l/mm grating was used, giving a wavelength coverage of 1.9
to 2.5~\micron~and the sources were observed with a single slit
position. A one pixel (0.6~arcsec) wide slit was used and the
detector was stepped over 2 pixels in half-pixel increments to
give a fully sampled spectrum with a spectral resolution of 600.
Observations were made through thick, patchy cirrus causing large
variations in the detected flux. Individual exposures were
weighted by the square of the signal to noise ratio of the
$v=1-0~\mathrm{S}(1)$ line before they were added together to
maximise the signal to noise ratio of the combined data. The
spectrum was ratioed by a standard star with its blackbody shape
corrected, to remove atmospheric telluric features. The final
spectra of the two sources are shown in Figure~3.

\subsection{Integral field spectroscopy}
\label{ifu_obs}

Sources A and B were observed on 24 October 2002 as part of the
UIST commissioning observations. UIST is a new facility class
near-IR (1--5~\micron) imager and spectrometer at UKIRT which uses
a $1024 \times 1024$ InSb array \citep{rh00}.  It includes a
deployable cryogenic image-slicing IFU with 14 slices, each
0.24~arcsec wide. The pixel scale along the slices is
0.12~arcsec/pixel resulting in a field of view of 3.36~arcsec
$\times$ 5.52~arcsec, which can be rotated to any angle on the
sky. The HK grism was used, giving a spectral coverage of
1.4--2.5~\micron\ with a spectral resolution of 800--1000.

The target was acquired using UIST in \textit{K}-band imaging mode.
The IFU field of view was rotated to a position angle of $90^\circ$,
making the long axis of the field run E-W. Two adjustments to the
pointing were made during the observations (1~arcsec E, then 1~arcsec
S). A total of three positions were observed for 8mins each giving a
maximum integration of 24mins on-source for the overlap region. Sky
frames were obtained by offsetting to a separate sky position. The
standard star BS7260 was observed to provide correction for telluric
atmospheric features and flux calibration.

\subsubsection{Combining the data into a single datacube}

The observations at each of the three positions described above were
formed into an $(x, y, \lambda)$ data-cube using the \textsc{orac-dr}
pipeline employed for all on-line data reduction at UKIRT. Each frame was
divided by a flat-field frame.  Sky-subtracted frames were wavelength calibrated using a
spectrum of an argon arc lamp and the resulting frames formed into a
datacube. All of the spectra in the datacube were divided by the
standard-star spectrum, corrected for the black-body shape of the
stellar continuum, to remove variations in atmospheric
transmission. An approximate flux calibration was obtained using the
magnitude of the star (K=4.48). The IFU data
reduction pipeline is described in more detail in \citet{todd02}.

The known 1~arcsec telescope offsets were used to register the
three data-cubes in the two spatial dimensions. These were
mosaiced into a single data-cube using \textsc{makemos}, which is
included in the Starlink \textsc{ccdpack} package. An image
extracted from this datacube at the wavelength of the
$v=1-0~\mathrm{S}(1)$ $\mathrm{H_2}$ line (2.122~\micron) is shown
in Figure~4. A spectrum of the sum of the brightest part of source
A is shown in Figure~5. All detected emission lines are from
$\mathrm{H_2}$.

\begin{figure}
  \centering
    \includegraphics[width=6cm]{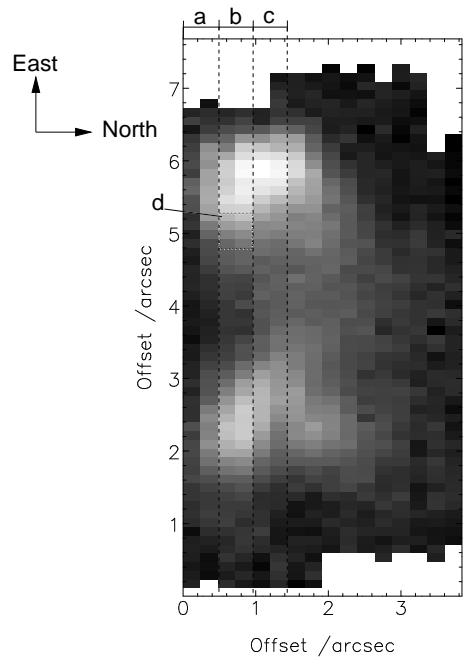}
  \caption{The spectral line image of sources A and B at the wavelength
  of the $1-0~\mathrm{S}(1)$ line, extracted from the IFU datacube. The
  offsets are arbitary and from the edge of the IFU field, which was
  centred on $\alpha=18^{\rm h}34^{\rm m}23.5^{\rm s},
\delta=-6^\circ53'00''$. At a distance of 3kpc, 1arcsec corresponds to
    0.015pc.}
\label{s1_image}
\end{figure}

\begin{figure*}
\begin{minipage}{140mm}
  \centering
  \includegraphics[width=14cm]{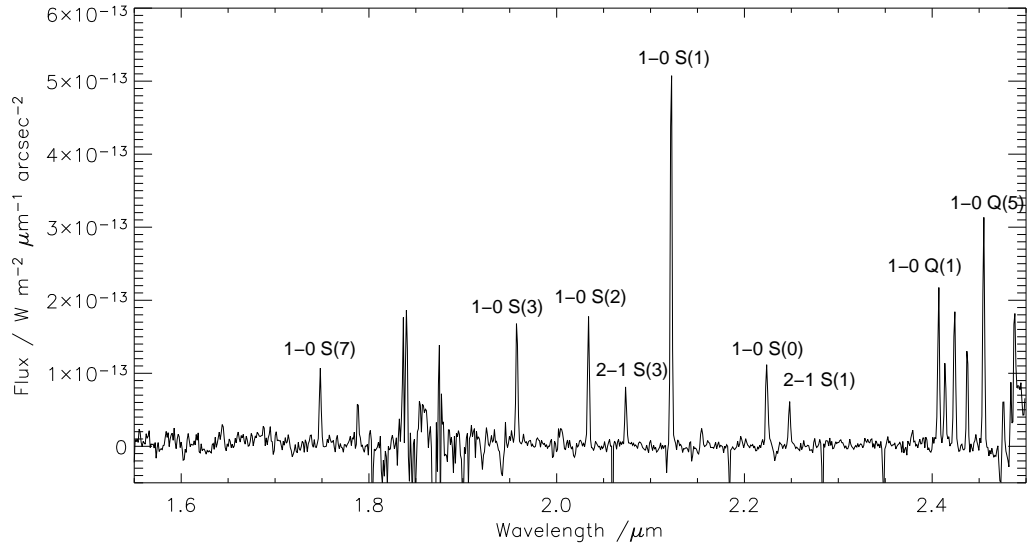}
  \caption{The spectrum obtained by summing over the brightest region of
  source A. Fluxes measured from this spectrum are given in Table~1.}
\label{bright_spec}
\end{minipage}
\end{figure*}

A spectrum was formed by summing over all spatial pixels of the
data-cube.  A Gaussian profile was fitted to the bright
$1-0~\mathrm{S}(1)$ line in this spectrum and was found to be
centred at $(2.12209 \pm 0.00003)$~\micron\ with FWHM $(2.28 \pm
0.05) \times 10^{-3}$~\micron. The other lines are all much
weaker, so to obtain the best possible measurement of the flux in
each line a Gaussian profile was scaled to fit the line keeping
the FWHM of the line fixed at 0.00228~\micron\ and the offset of
the line from the $1-0~\mathrm{S}(1)$ line fixed by the known
wavelength of the line. In total, 12 lines from the v=1 and v=2
transitions were detected. The measured fluxes are reported in
Table~1.

\begin{table*}
\begin{minipage}{110mm}
\begin{center}
\caption{Measure fluxes and physical parameters for the lines measured.}
\begin{tabular}{| l | r | r | c | r | r |}
\hline
Transition & $\lambda_j$ /\micron & $E_j$ /K & $A_j$ /$10^{-7}$~s &
\hspace{0.2cm}$g_j$ & Flux /$\mathrm{W~m^{-2}~arcsec^{-2}}$\\
\hline
$1-0~\mathrm{S}(7)$ & 1.7480 & 12817 & 2.98 & 57 & $(2.4 \pm 0.2)
\times 10^{-16}$\\
$1-0~\mathrm{S}(3)$ $^\dagger$ & 1.9576 & 8365 & 4.21 & 33 & \\
$1-0~\mathrm{S}(2)$ & 2.0338 & 7584 & 3.98 & 9 & $(4.4\pm 0.2)
\times 10^{-16}$\\
$2-1~\mathrm{S}(3)$ & 2.0735 & 13890 & 5.77 & 33 & $(1.7\pm 0.2)
\times 10^{-16}$ \\
$1-0~\mathrm{S}(1)$ & 2.1218 & 6956 & 3.47 & 21  & $(11.4 \pm 0.2)
\times 10^{-16}$ \\
$1-0~\mathrm{S}(0)$ & 2.2235 & 6471 & 2.53 & 5 & $(2.7\pm 0.2)
\times 10^{-16}$ \\
$2-1~\mathrm{S}(1)$ & 2.2477 & 12550 & 4.98 & 21 & $(1.7\pm 0.2)
\times 10^{-16}$ \\
$1-0~\mathrm{Q}(1)$ $^\dagger$ & 2.4066 & 6149 & 4.29 & 9 & \\
$1-0~\mathrm{Q}(2)$ $^\dagger$ & 2.4134 & 6471 & 3.03 & 5 & \\
$1-0~\mathrm{Q}(3)$ $^\dagger$ & 2.4237 & 6956 & 2.78 & 21 & \\
$1-0~\mathrm{Q}(4)$ & 2.4375 & 7586 & 2.65 & 9 & $(3.0 \pm 0.3)
\times 10^{-16}$ \\
$1-0~\mathrm{Q}(5)$ $^\dagger$ & 2.4548 & 8365 & 2.55 & 33  & \\
\hline
\end{tabular}
\end{center}
\medskip

The wavelengths, upper energy levels, Einstein $A$
  coefficients and degeneracies of the $\mathrm{H_2}$ lines detected
  in our spectra. Lines marked $^\dagger$ were found to be absorbed by
 unresolved atmospheric absorption features and were discarded. The
  fluxes given are measured from the spectrum in Figure~5 and are not corrected for extinction.

\label{h2_table}
\end{minipage}
\end{table*}

\section{Results and Discussion}

\subsection{Morphology}

The K--band image of G25.65+1.05 (Figure~1a) is dominated by the
nebulous continuum emission surrounding the central, unresolved,
source at $\alpha=18^{\rm h}34^{\rm m}20.9^{\rm s},
\delta=-6^\circ02'42.3''$ seen by Zavagno et al. (2002). The
morphology of the $\mathrm{H_2}$ emission in Figure~4 is strongly
suggestive of a moderately well-collimated outflow, or outflows,
centred on G25.65+1.05. Sources A, B, C and D are connected by
diffuse emission and appear to form one lobe of an outflow with a
flow axis through G25.65+1.05 to source E. The outflow has a
position angle on the sky of 130$^\circ$ East of North. This is
consistent with the direction of the highly energetic bipolar
outflow found to be centred on or close to the \textsc{H~ii}
region by \citet{shep2}. Sources ABCD are spatially co-incident
with the blueshifted lobe seen in the CO outflow and E with the
redshifted lobe. The relative faintness of E compared with ABCD
may then be interpreted as being due to E being embedded further
in the molecular cloud. The projected length of the outflow, taken
to be the distance from C to E, is $\sim$70arcsec. In the absence
of any information, we assume an angle relative to the plane of
the sky of 45$^\circ$. This gives an estimate of the total length
of 1.4pc and a collimation factor $\sim$3. Interpreted as a single
outflow, we find a lower degree of collimation than is seen in
outflows from low-mass YSOs, but higher than the 1-1.8 often
assumed for high-mass YSOs.

\citet{shep2} found two velocity components associated with
G25.65+1.05 at 39km${\rm s}^{-1}$ and at 49km${\rm s}^{-1}$. One
possible interpretation of the $\mathrm{H_2}$ emission in this
region would be to see sources A and B as bow-shocks in a highly
collimated flow from west to east, as suggested by their
morphology. This would imply the presence of another, more deeply
embedded source to the south and west of G25.65+1.05. The MSX
point source catalogue \citep{Egan98} shows a single source
associated with G25.65+1.05 ($\alpha=18^{\rm h}34^{\rm m}21.2^{\rm
s}, \delta=-6^\circ02'36.96''$ with a positional accuracy of
$4''-5''$) so does not support an argument for a second source
offset south-west of G25.65+1.05. Furthermore, a recent submillimetre
survey by \citet{hill05} finds a single core of mass
1.$\times10^3 \rm M_{\odot}$ associated with G25.65+1.05,
located at the position of the methanol
maser \citep{molinari96}, consistent with an earlier result by
\citet{faundez04}.

In the following Section, the near-IR spectra of sources ABCD are
analysed and shown to support our model of the emission arising due to
a single outflow.

\subsection{$\mathrm H_2$ Excitation in the G25.65+1.05 Outflow}

The near-infrared emission from $\mathrm{H_2}$ can be produced
either by thermal excitation in shock-fronts or fluorescent
excitation by non-ionizing ultraviolet photons ($91.2 < \lambda <
110.8$~nm) from hot young stars. These cases may be distinguished
using the near-IR spectrum, though the detection of lines from a
broad range of energy levels is required to obtain a secure
result. For shock-excited $\mathrm{H_2}$, the lower energy levels (v=1) are
typically populated as for a gas in local thermal equilibrium
(LTE) with a characteristic excitation temperature of a few
thousand Kelvin, though in reality the temperature of the cooling
gas will vary over a range from a few hundred to a few thousand
Kelvin. The emission lines would also be broadened to tens of
kms$^{-1}$. This is less than our instrumental resolution
and undetectable in the spectroscopy of G25.65+1.05.

For radiatively excited gas, the populations follow a non-LTE
distribution, characterised by high excitation temperatures
($\sim10,000$K). Lines from high vibrational levels (v$>>1$) may be detected and
the ratio of the intensity of the $1-0~\mathrm{S}(1)$ line to that
of the $2-1~\mathrm{S}(1)$  $\mathrm{H_2}$ line is $\sim 2$ if the
density is low \citep{sternberg89}.  For gas above the critical
density (${\rm n}_{\rm H_2}\sim 10^5~\mathrm{cm^{-3}}$),
collisional de-excitation becomes important and the level
populations tends towards an LTE population.

The spectra of G25.65+1.05 provide high signal/noise observations of
lines from the v=1 and v=2 vibrational levels. Deeper observations would
be required to obtain line strengths for the higher vibrational levels
and thereby to confirm the excitation. However, there are indicators
that the $\mathrm{H_2}$ is shock excited which, taken in aggregate, lead
us to conclude that this is indeed the case.

In the following sections, we obtain the excitation temperatures for the
sources in G25.65+1.05. These are in the range 1000-3000K indicative of
thermal excitation. When molecular hydrogen is radiatively excited,
90$\%$ of the $\mathrm{H_2}$ is dissociated by the ionizing photons
($\lambda < 91.2$~nm) that are also present. This may result in emission
in the 2.166~\micron\ Brackett-$\gamma$ recombination line. The spectra of
sources ABCD contain only emission lines of $\mathrm{H_2}$. We note that
the absence of Brackett-$\gamma$ emission
is not a strong constraint as some fluorescently excited
photodissociation regions emit only $\mathrm{H_2}$ lines. Finally we
cite the bow-shock morphology seen clearly in Figure~4, perhaps the
strongest indicator that the $\mathrm H_2$ is excited by a shock front
driven by an outflow rather than heated from within the knots by
embedded point sources.

\subsubsection{The Excitation Temperature of Sources A and B}

The measured intensity, $I$, of a given $\mathrm{H_2}$ line can be used to
calculate the column density of the upper excitation level of the
transition:
\begin{equation}
N_j = \frac{4 \pi \lambda_j I}{A_j hc}
\label{col_dens_eq}
\end{equation}
where $A_j$ is the Einstein $A$-coefficient of the transition. The
relative column densities of any two excitation levels can be
expressed in terms of an excitation temperature $T_{\mathrm{ex}}$:
\begin{equation}
\frac{N_i}{N_j} = \frac{g_i}{g_j} \exp \left(\frac{-(E_i - E_j)}{kT_{\mathrm{ex}}} \right)
\label{thermal_eq}
\end{equation}
where $g_j$ is the degeneracy and $E_j$ is the energy of the
level. The values of $\lambda_j$, $E_j$, $A_j$ and $g_j$ for the
lines detected in our spectra are shown in Table~1.

Before we could use our measured intensities to derive the excitation
temperature of the gas it was necessary to measure and compensate for
extinction.  In the absence of more information it was assumed that
the extinction was constant across the IFU field of view. An
extinction law of the form $\tau(\lambda) = A_k\left(\lambda/2.2
  \micron\right)^{-1.75}$ was used, giving a corrected intensity of
$I_\mathrm{corr} = I/e^{-\tau(\lambda)}$. Plotting $\log(N_j/g_j)$
against $E_j$ should give a straight line for low vibration levels and
for a single characteristic excitation temperature in each
pixel. If the value of $A_k$ used to correct the line intensities from which the column
density is calculated is wrong then the scatter of the points will
increase. Our measurements are particularly sensitive to this because
we have measured one \textit{H}-band line ($1-0~\mathrm{S}(7)$ at
1.748~\micron) which comes from an upper level close in energy to the
upper level of two \textit{K}-band lines ($2-1~\mathrm{S}(1)$ at
2.248~\micron\ and $2-1~\mathrm{S}(3)$ at 2.074~\micron).

The value of $A_k$ which minimised the scatter of the points was
measured for each spatial pixel in the two brightest columns in the
rebinned data-cube. It was found that the intensities of the
$1-0~\mathrm{S}(3)$ line and all the Q-branch lines other than the
$1-0~\mathrm{Q}(4)$ line were not consistent with any non-negative
values of $A_k$.  These lines were therefore assumed to be partly
absorbed by the very narrow atmospheric lines which dominate the edges
of the \textit{K}-band window which, being spectrally unresolved, are
not removed by dividing by the standard star. Once these lines were
excluded the mean value of $A_k$ over all the spatial pixels used was
found to be $A_k = 0.7 \pm 0.1$, where the error on the derived value
was estimated from the scatter of the values measured from one spatial
pixel to another. The variation in the measurements appeared random,
with no evidence for any systematic variation in extinction across the
source. There was also no evidence of curvature in the $N_j/g_j$ versus
$E_j$ plots, one of which is shown in Figure~6, though the absence of
any points with upper energy levels between 8000 and 12000~K would make
it hard to detect small deviations from a straight line. A curved line
in this plot would provide evidence for non-LTE processes, such as would
be seen for radiatively excited gas.

\begin{figure}
  \centering
    \includegraphics[width=8cm]{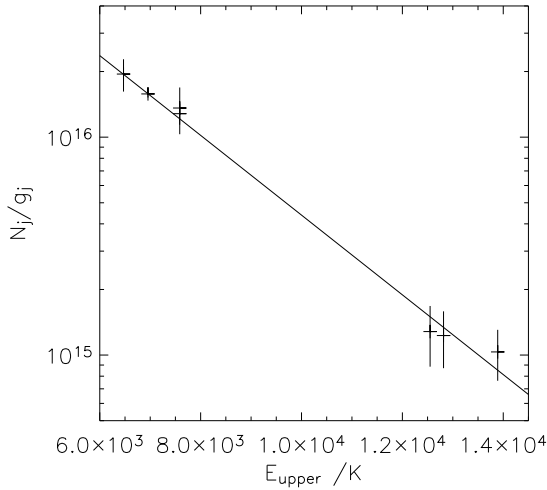}

  \caption{Plotting $N_j/g_j$ against $E_{\mathrm{upper}}$ on a logarithmic
    scale gives a straight line after correcting for extinction using
    the extinction law described in the text. The solid line is given
    by Equation~(3) using the maximum likelihood values
    of $T$ and $\alpha$. These measurements are taken from the
    0.48arcsec$\times$0.48 arcsec region of the source marked 'd' Figure~4.}
\label{col_dens_plot}
\end{figure}

The line fluxes were measured at each spatial pixel in the
rebinned data-cube and corrected for extinction using the
extinction law described above. The value of $N_j/g_j$ was then
derived from each corrected line flux using Equation~(1). The
excitation temperature was measured by fitting Equation~(2) to all
of the measurements made at a single spatial pixel using a maximum
likelihood method. When a single intensity is used there is a
degeneracy between $T$ and the constant of proportionality,
causing the most likely value of the constant of proportionality
to vary exponentially with $T$. For this reason Equation~(2) was
reformulated as
\begin{equation}
\frac{N_j}{g_j} = \exp \left(\frac{-E_j}{kT} + \alpha \right)
\label{thermal_eq2}
\end{equation}
where $\alpha$ is the logarithm of the constant of proportionality.

The likelihood of a range of values of $T$ and $\alpha$ was calculated
for each measured line flux assuming a gaussian error distribution on
the measurements of the line intensity and using a uniform prior
probability density function for both parameters:
\begin{equation}
L_j(T, \alpha) \propto  \frac{\exp[{-(\delta/2\sigma)^2}]}{\sigma \sqrt{2\pi}}
\label{likelihood}.
\end{equation}

In Equation~(4) $\delta$ is the difference between the value of
$N_j/g_j$ derived from the measured intensity using Equation~(1)
and that calculated using Equation~(3) and $\sigma$ is the
$1\sigma$ error on the measured value of $N_j/g_j$.  When this is
evaluated for a single spectral line there is a degeneracy between
the two parameters, as shown in Figure~7a. The likelihood of the
parameters using all the line fluxes is calculated by taking the
product of all the individual likelihoods:
\begin{equation}
L(T, \alpha) = \prod_j{L_j(T, \alpha)}
\end{equation}
This reduces the degeneracy (Figure~7b).  The likelihood of each
value of $T$, taking into account the uncertainty in $\alpha$, can
be calculated by marginalising the likelihood, or summing over all
values of $\alpha$ to produce a one dimensional likelihood curve
(Figure~8). The 68\% confidence levels (the narrowest possible
range of values of $T$ to include 68\% of the total area under the
likelihood curve) were measured as an equivalent of $1\sigma$
error bars on a gaussian distribution.

\begin{figure}
  \centering
    \includegraphics[width=7cm]{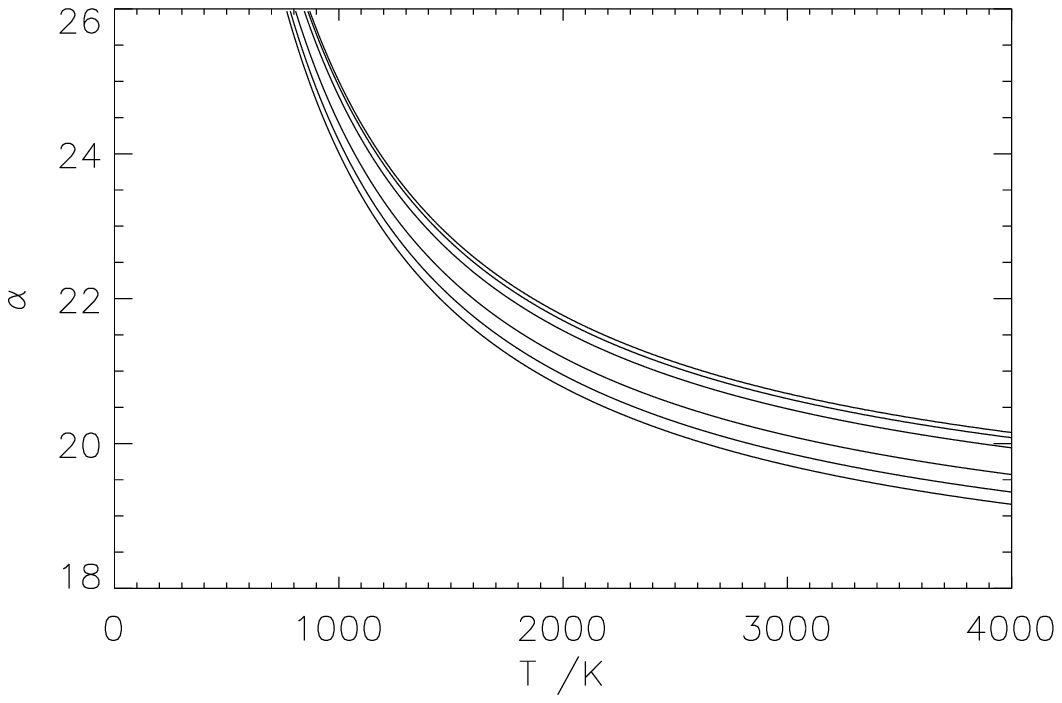}
    \includegraphics[width=7cm]{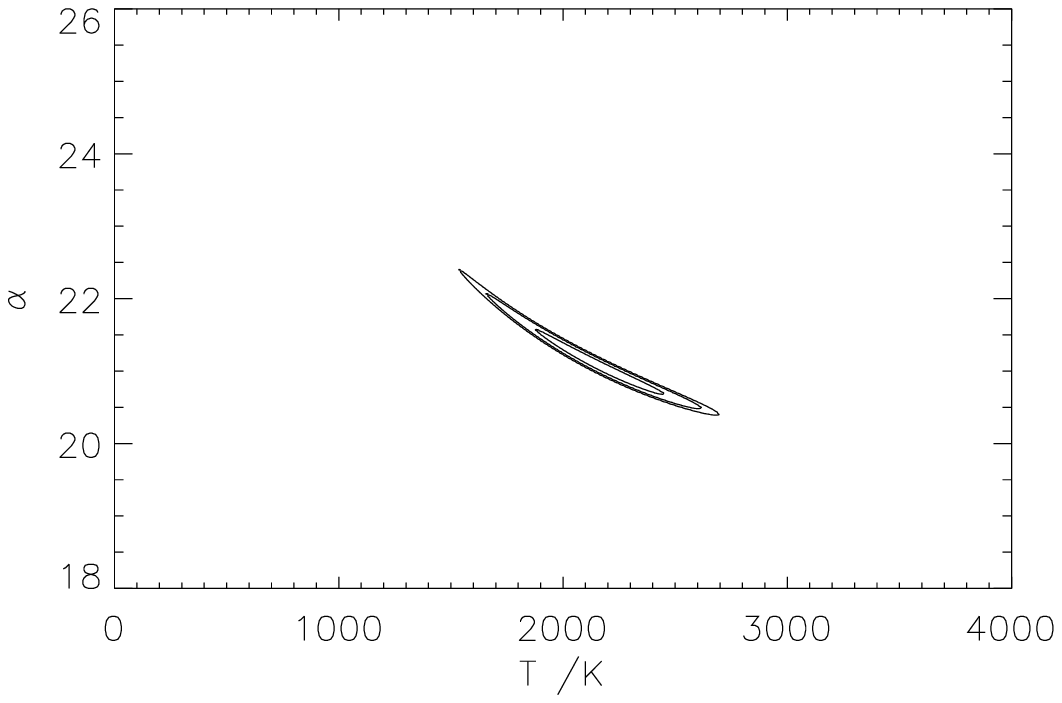}
  \caption{(a) The likelihood of the parameters $T$ and $\alpha$ were
    evaluated for a single spatial pixel using the intensity of a
    single spectral line. (b) When other spectral lines are introduced
    the degeneracy is partly broken.  On both plots the contours are
    at the 68\%, 95\% and 99\% confidence levels, equivalent to
    $1\sigma$, $2\sigma$ and $3\sigma$ error bars on a gaussian
    distribution.}
\label{likelihood_1}
\end{figure}

\begin{figure}
  \centering
 \includegraphics[width=7cm]{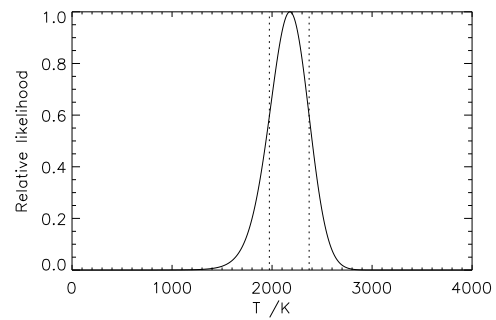}

  \caption{The marginal distribution of $T$ can be obtained by
    integrating the distribution shown in Figure~7b
    over $\alpha$. The likelihood shown here is normalised to have a
    maximum value of 1. The maximum likelihood value of $T$
    (2178~K in this case) and confidence intervals can be measured
    from this. The 68\% confidence interval of 1973 -- 2371~K is shown
    by the dotted lines.}
\label{likelihood_2}
\end{figure}

These measurements were carried out along three 0.48~arcsec wide
strips marked a, b and c on the image shown in Figure~4. The results are
shown in Figure~9.  The excitation temperature clearly increases from around
1800~K to $(2840 \pm 230)$~K (averaging over the most easterly
pixel of all three slices) at what may be the shock front. In
source A there also appears to be a decrease in excitation temperature of
$\sim 200$~K to the right (east) in the image shown in Figure~4.

\begin{figure}
  \centering
   \includegraphics[width=7cm]{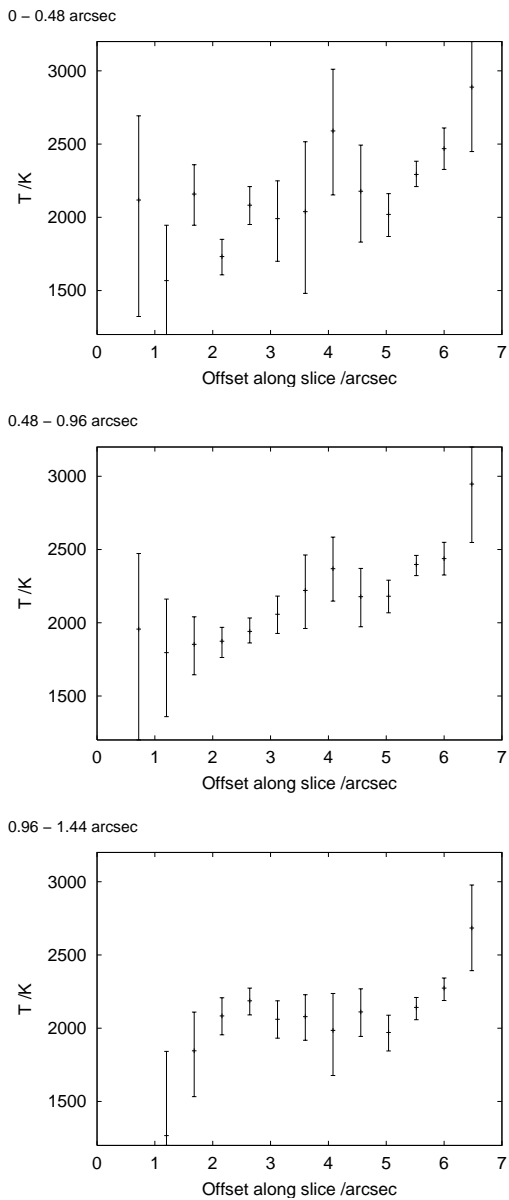}\\
  \caption{The temperature was measured along three 0.48~arcsec wide
    strips running west-east on the left of Figure~4.
    The points and error bars on these plots show the maximum
    likelihood value and 68\% confidence interval on the value of $T$
    marginalised over $\alpha$, as described in the text.}
\label{temperatures}
\end{figure}

In Figure~4 it can be seen that both sources are brighter on the
left (south) side than on the right. The temperatures measured in
source A are approximately 200~K higher on the south edge of the
IFU field than 1.5~arcsec to the north. We interpret this
asymmetry in the north-south direction as supporting our
interpretation of the $\mathrm{H_2}$ emission being associated
with a single outflow. If sources A and B are were produced by a
flow from west to east a symmetrical bow-shock would be expected.
However, we cannot exclude the possibility that the asymmetry is
produced by variations in density in the medium with which the
outflow is interacting.

\subsubsection{The Excitation Temperature of Sources C and D}

The spectra of sources C and D are shown in Figure~3 and the
relative line intensities in Table~2. Although the weather was
non-photometric, both sources were observed simultaneously and the
line ratios may be considered to be accurate, at the limit of the
gaussian errors. The maximum likelihood analysis was repeated for
C and D, giving an excitation temperature of 2730~K with 68\% confidence
levels of (2727$\pm$227)K for source C and (2256$\pm$250)K for source D,
in each case averaged over the observed extent of the source.

\begin{table*}
\begin{minipage}{60mm}
\begin{center}
\caption{Line ratios from the spectra of sources C and D.}
\begin{tabular}{| l | c | c |}
\hline
Line &  \multicolumn{2}{c}{Ratio to 1-0 S(1)} \\
     &  Source C  & Source D \\
\hline
1-0 S(1) & 1 & 1 \\
2-1 S(1) & 0.15 $\pm$ 0.02  & 0.10 $\pm$ 0.04 \\
1-0 S(0) & 0.20 $\pm$ 0.05  & 0.36 $\pm$ 0.07 \\
2-1 S(3) & 0.22 $\pm$ 0.04 & 0.15 $\pm$ 0.05 \\
1-0 S(2) & 0.28 $\pm$ 0.08  & 0.52 $\pm$ 0.08 \\

\hline
\end{tabular}
\end{center}
\medskip

The ratio of line fluxes measured in the CGS4 spectra of sources C and
D relative to the 1-0 S(1) line flux. The spectra were observed in
cloudy conditions, so the absolute fluxes are unknown.

\label{cgs4tab}
\end{minipage}
\end{table*}

\subsubsection{Shock excitation}

The properties of four classes of shocks were summarised by
\citet{davies00}:
\begin{enumerate}
\item \textit{fast J shock} (100--300~$\mathrm{km~s^{-1}}$): hydrogen
  molecules are dissociated and reform producing a spectrum similar to
  that of UV fluorescence. The flux of Brackett-$\gamma$ and $H$-band
  Fe~\textsc{ii} lines are comparable to the $1-0~\mathrm{S}(1)$
  flux.

\item \textit{slow J shock}: strong $\mathrm{H_2}$ lines are produced
  and molecules are not dissociated. However, in normal ionization
  fractions, magnetic field strengths and gas densities, such shocks
  are expected to evolve into C-type shocks.

\item \textit{fast C shock}: shock velocities of $\sim
  40~\mathrm{km~s^{-1}}$ heat the gas to $\sim 2000$~K or more (dependent
  on the velocity) producing strong $\mathrm{H_2}$ lines.

\item \textit{slow C shock}: peak temperature 300~K or less, resulting
  in very weak $\mathrm{H_2}$  emission.

\end{enumerate}

The absence of Brackett-$\gamma$ emission from atomic hydrogen or
Fe~\textsc{ii} lines and the excitation temperatures measured in all
the sources ABCD leads us to suggest that the fast C shock model may be most
appropriate. A maximum post-shock excitation temperature of $(2840 \pm 230)$~K
would be produced by a shock velocity of
$(37\pm3)~\mathrm{km\:s^{-1}}$ in the models of \citet{kaufman96}.
This would be consistent with the $29~\mathrm{km\:s^{-1}}$ along the
line of sight measured by \citet{shep2} when the uncertainty on
the angle of incidence is taken into account. The temperatures
measured in source A (offsets along the slice of between 4.5 and
6.5~arcsec) decrease with distance from the shock front.  The
C-shock models of \citet{flower96} predict a steady decrease in
temperature over around 0.03~pc (2~arcsec at a distance of 3~kpc)
behind the shock front, beyond which point the flux falls
rapidly. This is consistent with what we see, though the proximity of
source B and the uncertain geometry of the outflow inhibits more
detailed comparison with the models. A more detailed analysis, including
lines from higher vibrational levels examined as function of distance
across the shock-front or high spectral resolution to explore the lines
shapes would be required to allow us to draw stronger
conclusions. Integral field spectroscopy is a technique that would lend
itself readily to these observations.

\section{Summary}

Mapping at high spatial resolution in the near-IR emission from
$\mathrm{H_2}$ we have directly observed outflow activity from one or
more sources associated with G25.65+1.05. Long slit and integral field
spectroscopy of the brightest sources show that they are excited by a
shock. The measured excitation temperatures and absence of dissociation
lead us to suggest that a C-shock with a shock velocity of $(37 \pm
3)~\mathrm{km\:s^{-1}}$ may be responsible. Integral field spectra of a
region containing two compact sources of $\mathrm{H_2}$ emission showed
a variation in excitation temperature with a maximum of 2800~K close to
the shock front. The excitation temperature in the brighter of the two
sources falls steadily over around 0.03~pc at which point the intensity
decreases significantly, in agreement with the C-shock models of
\citet{flower96}.

Both the location of the brightest $\mathrm{H_2}$ emission and the
derived shock speed are consistent with the excited $\mathrm{H_2}$
tracing the same outflow as the CO emission detected by
\citet{shep2}.

We conclude that the most likely interpretation of these data is that
the $\mathrm{H_2}$ emission traces the full width of the outflow. In
this case, the collimation factor of the outflow is $\sim 3$, higher
than the 1 to 1.8 often though to be typical of outflows from high
mass stars.

The combination of near-infrared imaging and spectroscopy is a
powerful method of revealing outflows in regions of high mass star
formation. However, in the case of G25.65+1.05, the near-infrared
data alone cannot determine unambiguously whether there are
multiple or single outflows and driving sources. Supporting
information may be sought in a number of ways. Higher resolution
spectroscopy can help associated the emission with different
outflows (see e.g. \cite{Davis04} on IRAS18151-1208). High
resolution imaging at wavelengths from 3-20microns and longer,
such as is now possible from large ground based telescopes, may
reveal additional deeply embedded sources.

The combinination of high spatial resolution interferometric radio
mapping with infrared imaging has been used successfully by
e.g. \citet{lee01}, \citet{walsh02} and \citet{beuther03} to examine the detailed
structure and morphology of crowded regions containing multiple sources and outflows.
G323.74-0.26 is revealed by the observations of \citet{walsh02} to
contain both fluorescent ${\mathrm H_2}$ emission from the H{\sc ii}
region associated with this source and shocked ${\mathrm H_2}$ emission
from an outflow. Combining interferometric observations with infrared images of
IRAS 19410+2336 \cite{beuther03} found that what had appeared in single dish
maps to be two outflows was the combination of at least seven and
possibly as many as nine separate outflows. It seems likely that
similar results might be found in this region.

\section*{Acknowledgments}

The United Kingdom Infrared Telescope is operated by the Joint Astronomy
Centre on behalf of the U.K. Particle Physics and Astronomy Research
Council. The authors would like to thank the staff of UKIRT for their
support in carrying out the observations presented here. We are
grateful to the referee, Michael Burton, for his detailed comments on
the paper which have allowed us to improve it from the first submission.

\label{lastpage}

\end{document}